\documentclass[showpacs,twocolumn,aps,floatfix,preprintnumbers]{revtex4}

\usepackage{amsmath,amssymb}
\usepackage{graphicx}
\usepackage{dcolumn}
\usepackage{bm}
\begin{document}
%\preprint{Progress in Muon Research}
\title{The Lifetime Asymmetry of Polarized Muons in Flight}

\author{Zhi-Qiang Shi}\email{zqshi@snnu.edu.cn}
        %\altaffiliation[Permanent address: ]{Department of Physics, Shaanxi Normal University,
        %Xi'an 710062, China}

        %\affiliation{Faculty of Science, Xi'an Jiaotong University, Xi'an 710049, China}
        \affiliation{Department of Physics, Shaanxi Normal University, Xi'an 710062, China}

\author{Guang-Jiong Ni}
        \email{pdx01018@pdx.edu}
        \affiliation{Department of Physics, Fudan University, Shanghai, 200433, China\\
        Department of Physics, Portland State University, Portland, OR 97207 USA}

%\date{}
\begin{abstract}
Based on the parity violation in Standard Model, we study the dependence of the lifetime
on the polarization of an initial-state fermion in weak interactions. The fermion
lifetime was usually calculated in terms of spin states. However, after comparing spin
states with helicity states and chirality states, it is pointed out that a spin state is
helicity degenerate, and the spin state and the helicity state are entirely different.
Using helicity states, we calculate the lifetime of polarized muons. The result shows
that the lifetime of right-handed polarized muons is always greater than that of
left-handed polarized muons with the same speed in flight.
\end{abstract}

\maketitle

\section{introduction}

In virtue of parity violation the experiments have shown that all fermions emitted from
decay processes are longitudinally polarized. It is well known that neutrinos are
left-handed (LH) polarized while antineutrinos are right-handed (RH) polarized. In beta
or muon decays it is found that electrons are LH polarized while positrons are RH
polarized.$^{[1,2]}$ Based on these experimental facts, naturally, it is thought that not
only the final-state fermions, but also the initial-state fermion should reveal the
feature of longitudinal polarization in weak interactions. So we could further consider
that the lifetime of fermions in the LH helicity state and that in the RH helicity state
should be different. Namely, the lifetime asymmetry should exist in left-right handed
polarized fermions. However, the experimental and theoretical study on the polarization
or helicity of the initial fermions in decays has not yet been discussed fully in the
literature. In a previous paper$^{[3]}$ we have given a first discussion of the problem
on lifetime asymmetry. In this paper we continue the study of this issue. The differences
among spin states, helicity states and chirality states are emphasized and the lifetime
asymmetry of polarized muons will be calculated concretely.

\section{the charged weak currents \protect\\ in the standard model}
In the standard model (SM), all of fundamental fermions are divided into two classes, LH
chirality state and RH chirality state, in order to describe the parity violation in weak
interactions. The LH chirality state is different from the RH chirality state. The former
is a SU(2)-doublet state whereas the latter a SU(2)-singlet state, hence they have
different gauge transformations. Especially, RH chirality states have zero weak isospin
and are only present in neutral weak currents. Therefore, only LH chirality states exist
in charged weak currents. The interaction Lagrangians for charged weak lepton current and
charged weak quark current read, respectively
\begin{eqnarray}
    {\cal L}_{\ell\:W}&=& \frac{1}{\sqrt{2}}\:g_{_2}\:{\overline e_{_L}}\gamma_\mu
    W_\mu^+\nu_{_L} +\frac{1}{\sqrt{2}}\:g_{_2}\:{\overline \nu_{_L}}\gamma_\mu
    W_\mu^-e_{_L},\label{eq:1}\\
    {\cal L}_{q W}&=& \frac{1}{\sqrt{2}}\:g_{_2}\:{\overline d_{_L}}\gamma_\mu W_\mu^+u_{_L}
    +\frac{1}{\sqrt{2}}\:g_{_2}\:{\overline u_{_L}}\gamma_\mu
    W_\mu^-d_{_L},\label{eq:2}
\end{eqnarray}
where $g_{_2}$ is the coupling constant corresponding to SU(2) and the subscript $L$
denotes the LH chirality state. Obviously, all fermions are in the LH chirality states
while all antifermions are in the RH chirality states.

For example, the weak interaction in muon decay is successfully described by four-fermion
interaction Hamiltonian. We denote the matrix element by
\begin{equation}\label{eq:3}
    M\sim\sum g^\gamma_{_{\varepsilon\mu}}\langle\overline e_\varepsilon|\Gamma_\gamma|
    (\nu_e)_n\rangle\langle(\overline\nu_\mu)_m|\Gamma_\gamma|\mu_\mu\rangle,
\end{equation}
where $\gamma=S,V,T$ indicates a scalar, vector or tensor interaction, and $\varepsilon,
\mu=R,L$ indicate a right- or left-handed chirality of the electron or muon. The
chiralities $n$ and $m$ of the $\nu_e$ and $\overline\nu_\mu$ are then determined by the
values of $\gamma, \varepsilon$ and $\mu$. All coupling constants have been obtained
entirely from experiments without any model assumption.$^{[2,4]}$ The experiments on muon
decay show $g_{_{RL}}$, $g_{_{RR}}$, $g_{_{LR}}$ to be zero, and at least one of the two
coupling, $g^V_{_{LL}}$ or $g^S_{_{LL}}$, to be nonzero. The experiments on inverse muon
decay provide a lower limit for pure $V-A$ interaction with $|g^V_{_{LL}}|>0.960$. Thus
the measurements give a strong support to the SM which sets $g^V_{_{LL}}=1$ while all
others being zero, and then indicate that the charged weak current is dominated by a
coupling to left-handed chirality fermions. Therefore, the negative muon decay can be
written as
\begin{equation}\label{eq:4}
    \mu_{_L}^-\longrightarrow e_{_L}^{-}+\overline\nu_{e{_R}}+\nu_{\mu{_L}}.
\end{equation}
And the matrix element (3) has the form
\begin{equation}\label{eq:5}
    M\sim(\overline\nu_{\mu_L}\gamma_\mu\:\mu_{_L})(\overline e_{_L}\gamma_\mu\:\nu_{e_L}).
\end{equation}

\section{spin states, helicity states and chirality states}
There exist three kinds of spinor wave functions, i.e., the spin states, the helicity
states and the chirality states. We will discuss the difference and the relation among
them.

\subsection{The spin states}

The spin states are the plane wave solutions of Dirac equation, and in momentum
representation for a given four-momentum $p$ and mass $m$, the positive energy solution
and the so-called negative energy solution are respectively
\begin{equation}\label{eq:6}
  u_s(p)=\sqrt{\frac{E_p+m}{2E_p}}\left(\!\begin{array}{c}\varphi_s\\
  \frac{\displaystyle \bm \sigma\cdot\bm p}{\displaystyle E_p+m}\varphi_s\;
  \end{array}\!\right)\;,
\end{equation}
\begin{equation}\label{eq:7}
  v_s(p)=\sqrt{\frac{E_p+m}{2E_p}}\left(\!\begin{array}{c}\frac{\displaystyle \bm \sigma\cdot\bm p}
  {\displaystyle E_p+m}\varphi_s\\\varphi_s
  \end{array}\!\right)\;,
\end{equation}
where $E_p>0,\; s=1, 2$ and $\varphi_s$ are Pauli spin wave functions. The state with
$s=1$ is spin up ($\sigma_z=1$) while the state with $s=2$ is spin down ($\sigma_z=-1$).
They are eigenstates of operator, $\frac{\omega(\bm p)\cdot e}{m}$, with eigenvalues $\pm
1$, namely
\begin{equation}\label{eq:8}
  \frac{\omega(p)\cdot e}{m}\;u_s(p)=\left\{\begin{array}{lr}u_s(p),\quad&(s=1)\\
  -u_s(p).\quad&(s=2)\end{array}\right.
\end{equation}
Here $\omega(p)$ is the Pauli-Lubanski covariant spin vector and $e$ is the
four-polarization vector in the form
\begin{equation}\label{eq:9}
  e_{\alpha}=\left\{\begin{array}{ll}\bm e^0+\frac{\displaystyle\bm p\;(\bm p\cdot\bm e^0)}
  {\displaystyle m(E_p+m)}\;,\quad &(\alpha=1,2,3)\\
  i\;\frac{\displaystyle\bm p\cdot\bm e^0}{\displaystyle m}\;,\quad &(\alpha=4)
  \end{array}\right.
\end{equation}
which is normalized $(e^2=1)$, orthogonal to $p\;(e\cdot p=0)$. In the rest frame $e$
reduces to $e^0=(\bm e^0,0)=(0,0,1,0)$. In application it is frequently necessary to
evaluate spin sums in the form
\begin{equation}\label{eq:10}
  \begin{array}{l}
  P_1(p)=\rho_+\;\Lambda_+(p),\quad P_2(p)=\rho_-\;\Lambda_+(p),\\
  P_3(p)=\rho_+\;\Lambda_-(p),\quad P_4(p)=\rho_-\;\Lambda_-(p).
  \end{array}
\end{equation}
The $\Lambda_+(p)$ and $\Lambda_-(p)$ are the positive energy projection operator and the
negative energy projection operator,
\begin{equation}\label{eq:11}
  \Lambda_+(p)=\frac{-i\;\gamma\cdot p+m}{2E_p},\quad\Lambda_-(p)=\frac{i\;\gamma\cdot p+m}{2E_p},
\end{equation}
respectively and $\rho_{\pm}$ are spin projection operators:
\begin{equation}\label{eq:12}
  \rho_{\pm}=\frac{1}{2}(1\pm i\;\gamma_5\;\gamma\cdot e).
\end{equation}
The plus sign refers to $s=1$ and the minus sign to $s=2$. One sees that the operator
$P_1(p)$ project out the positive energy states with spin up and $P_2(p)$ the positive
energy states with spin down, whereas the operator $P_3(p)$ the negative energy states
with spin down and $P_4(p)$ the negative energy states with spin up in its rest frame.

\subsection{The helicity states}

A helicity state is the eigenstate of the helicity of fermions and satisfies the ordinary
Dirac equation$^{[5]}$. If spinor $\varphi$ is taken as the eigenstate of the spin
component along the direction of its motion,
\begin{equation}\label{eq:13}
  \frac{\bm \sigma\cdot
  \bm p}{|\bm p|}\;\varphi_{_h}=h\;\varphi_{_h},\quad h=\pm 1
\end{equation}
then the helicity states read
\begin{equation}\label{eq:14}
  u_{_h}(p)=\sqrt{\frac{E_p+m}{2E_p}}\left(\!\begin{array}{c}\varphi_{_h}\\
  \frac{\displaystyle h|\bm p|}{\displaystyle E_p+m}\varphi_{_h}
  \end{array}\right),
\end{equation}
and
\begin{equation}
    \varphi_{_{+1}}=\left(\!\begin{array}{c}\cos\frac{\theta}{2}\;e^{-i\frac{\phi}{2}}\\
    \sin\frac{\theta}{2}\;e^{i\frac{\phi}{2}}
    \end{array}\right),\;
    \varphi_{_{-1}}=\left(\!\begin{array}{c}-\sin\frac{\theta}{2}\;e^{-i\frac{\phi}{2}}\\
    +\cos\frac{\theta}{2}\;e^{i\frac{\phi}{2}}
    \end{array}\right),
\end{equation}
where $\theta$ and $\phi$ are the polar angles of momentum $\bm p$ in polar-coordinates.
The state with $h=+1$ is the RH helicity state while the state with $h=-1$ is the LH
helicity state. Because a helicity state is also a plane wave solutions of Dirac
equation, the energy projection operator of helicity state is equal to that of spin
state, namely
\begin{equation}\label{eq:15-1}
  \sum_{h}u_h(p)\overline{u}_h(p)=\sum_{s=1}^2u_s(p)\overline{u}_s(p).
\end{equation}
The projection operators of the helicity states are$^{[6]}$
\begin{equation}\label{eq:15}
  \rho_{_h}=\frac{1}{2}\left(1\pm \frac{\displaystyle \bm \Sigma\cdot \bm p}
  {\displaystyle |\bm p|}\;\right)\;.
\end{equation}

We can see that the helicity state is entirely different from the spin state. Both the
helicity eigenvalue and the projection operator of helicity state are not Lorentz
invariant, and the latter is essentially a two-component operator. From
Eqs.~(\ref{eq:6}), (\ref{eq:7}) and (\ref{eq:8}) we can find out that a spin state with
the same $s$ but different values of $h$ is helicity degenerate. So the spin states can
not uniquely describe the helicity of fermions. On the other hand, taking the simplest
case of $\bm p: \bm p=p_z$, which does not lose the universality of problem, we have the
LH helicity state $u_{_{Lh}}$ and the RH helicity state $u_{_{Rh}}$, respectively
\begin{eqnarray}
  u_{_{Lh}}(p)&=&\sqrt{\frac{E_p+m}{2E_p}}\left(\!\begin{array}{c}\varphi_2\\
  \frac{\displaystyle {-|\bm p|\;\varphi_2}}{\displaystyle {E_p+m}}\end{array}\!\right),
  \label{eq:16}\\
  u_{_{Rh}}(p)&=&\sqrt{\frac{E_p+m}{2E_p}}\left(\!\begin{array}{c}\varphi_1\\
  \frac{\displaystyle {|\bm p|\;\varphi_1}}{\displaystyle {E_p+m}}
  \end{array}\!\right).\label{eq:17}
\end{eqnarray}
Even so comparing Eq.~(\ref{eq:12}) with Eq.~(\ref{eq:15}) one can also see that the
projection operator of spin state is different from that of helicity state though the
spin and helicity state are formally identical when $\bm p=p_z$.

Furthermore, we should point out emphatically that the helicity and the degree of
polarization are two different concepts. The helicity $h$ describes the mutual relation
between the spin direction of fermions and its momentum direction. Because helicity
changes its sign under the space inversion, it is called a pseudoscalar. The degree of
polarization is the length of polarization vector $\bm P$ which is the ensemble average
of spin vector of fermion beam. The polarization vector does not change its sign under
the space inversion and so is called a pseudovector or an axial vector, like angular
momentum. The helicity can only take the value of either $1$ or $-1$ and has no meaning
in its rest frame. But the degree of polarization can take any value between $0$ and $1$
and the value remains constant in different frame. In the spin states the polarization of
fermions is described by polarization vector. It is the discrepancy between the helicity
and the polarization vector that results in the spin state being different from the
helicity state.

The four-polarization vector $e$ is the relativistic generalization of three-polarization
vector $\bm P$. It is able to prove from Eq.~(\ref{eq:12}) that the vector $\bm e$ is the
ensemble average of spin vector ${\bm \sigma}=-i\gamma_4\gamma_\mu\gamma_5$, i.e.,
\begin{equation}
    \bm e=\langle{\bm\sigma}\rangle=\bm P.
\end{equation}
On the other hand, however, we can see out from Eq.~(\ref{eq:9}) that the property of
vector $\bm e$ is different from that of pseudovector $\bm P$. For example, when $\bm
p=p_z$ vector $\bm e$ reads
\begin{equation}
    \bm e=\frac{E}{m}\bm e^0.
\end{equation}
Differing from vector $\bm P$, obviously, the direction of vector $\bm e$ is always
pointing to $z$ axis and its value can be greater than one, i.e., $|\bm e|\geq 1$.
Strictly speaking, only in the rest frame can the four-polarization vector $e$ be most
unambiguously defined$^{[7]}$. And the spin projection operators $\rho_{\pm}$, which are
Lorentz invariant, can only project out the states which in its rest frame have spin
$s=1$ and $2$, respectively. Therefore, the Pauli-Lubanski covariant spin vector
$\omega(p)$ and the four-polarization vector $e$, not like four-momenta, have no
intuitively and definitely physical significance in the motion frame. We reach a
conclusion that the polarization of fermions must be described by the helicity states
which are closely related to directly observable quantity experimentally.

\subsection{The chirality states}

The chirality states are the eigenstates of chirality operator $\gamma_5$. The LH
chirality state and the RH chirality state are defined as, respectively
\begin{equation}\label{eq:18}
  u_{_{LS}}(p)=\!\frac{1}{2}(1+\gamma_5)u_s(p),\quad
  u_{_{RS}}(p)=\!\frac{1}{2}(1-\gamma_5)u_s(p).
\end{equation}
In general, chirality states are different from helicity states. Only if $m=0$ (for
example neutrinos) or $E\gg m$ (in the ultrarelativistic limit) the fermions satisfy Weyl
equation$^{[3,8]}$, the spinor $\varphi_s$ must then be taken to be eigenstates of
helicity operator $h$ and the polarization is always in the direction of motion$^{[9]}$.
In other words, for $m=0$ the helicity states, the chirality states and spin states are
identical, i.e.
\begin{eqnarray}
  u^W_{Lh}(p)&=&u^W_{L2}(p)=u^W_2(p)=\frac{1}{\sqrt{2}}\left(\begin{array}{c}
  \varphi_2\\-\varphi_2\end{array}\right),\label{eq:19}\\
  u^W_{Rh}(p)&=&u^W_{R1}(p)=u^W_1(p)=\frac{1}{\sqrt{2}}\left(\begin{array}{c}
  \varphi_1\\\varphi_1\end{array}\right).\label{eq:20}
\end{eqnarray}
The superscript $W$ refers to it being a solution of Weyl equation.

The helicity states $u_{_{Lh}}$ and $u_{_{Rh}}$ in Eqs.~(\ref{eq:16}) and (\ref{eq:17})
can be expanded as linear combination of chirality states, respectively$^{[10]}$
\begin{eqnarray}
    u_{_{Lh}}(p)&=&\frac{1}{2}(1+\gamma_5)\;u_{_{Lh}}(p)+\frac{1}{2}(1-\gamma_5)\;u_{_{Lh}}(p)\nonumber\\
    &=&C_{LL}\;u^0_{_{L2}}+C_{LR}\;u^0_{_{R2}},\label{eq:21}\\
    u_{_{Rh}}(p)&=&C_{RL}\;u^0_{_{L1}}+C_{RR}\;u^0_{_{R1}},\label{eq:22}
\end{eqnarray}
where $u^0_{_{LS}}$ and $u^0_{_{RS}}$ are chirality states in the rest frame,
\begin{equation}\label{eq:23}
  u^0_{_{LS}}=\frac{1}{2}\left(\!\begin{array}{c}\varphi_s\\-\varphi_s
  \end{array}\!\right),\quad
  u^0_{_{RS}}=\frac{1}{2}\left(\!\begin{array}{c}\varphi_s\\\varphi_s
  \end{array}\!\right).
\end{equation}
The coefficients $C_{LL}$, $C_{LR}$, $C_{RL}$ and $C_{RR}$ as given by
\begin{eqnarray}
    C_{LL}=C_{RR}&=&\frac{1}{\sqrt{2E_p}}(\sqrt{E_p+m}+\sqrt{E_p-m})\nonumber\\
    &=&\sqrt{1+\beta},\label{eq:24}\\
    C_{RL}=C_{LR}&=&\frac{1}{\sqrt{2E_p}}(\sqrt{E_p+m}-\sqrt{E_p-m})\nonumber\\
    &=&\sqrt{1-\beta},\label{eq:25}
\end{eqnarray}
where $\beta$ is the velocity of the muons. It is obvious from Eqs.~(\ref{eq:21}) and
(\ref{eq:22}) that in a LH helicity state $u_{_{Lh}}(p)$ the coefficient $C_{LL}$ is the
amplitude of LH chirality state $u^0_{_{L2}}$ and the $C_{LR}$ that of RH chirality state
$u^0_{_{R2}}$; while in a RH helicity state $u_{_{Rh}}(p)$ the $C_{RL}$ that of LH
chirality state $u^0_{_{L1}}$ and the $C_{RR}$ that of RH chirality state $u^0_{_{R1}}$
in its rest frame. For LH helicity state the hidden amplitude of RH chirality state
decreases with the increase of $\beta$ until $C_{LR}\rightarrow 0$ when $\beta\rightarrow
1$, showing that a high-energy fermion can be LH polarized without hidden RH spinning
instability. For RH helicity state the hidden amplitude of LH chirality state decreases
with the increase of $\beta$. When $\beta=0$, the hidden amplitude of LH chirality state
is equal to that of RH chirality state and one even can not discriminate a rest fermion
being either LH or RH polarized$^{[11]}$.

\section{the lifetime of polarized muons}

Now let us consider a $\mu^-$ decay process (4). The lowest order decay rate or
lifetime $\tau$ for muon decays, based on the perturbation theory of weak interactions,
is given by
\begin{equation}\label{eq:26}
  \tau^{-1}=\frac{1}{(2\pi)^5}\!\int\!d^3 q\;d^3 k\;d^3 k'\;\delta^4(p-q-k-k')\;M^2.
\end{equation}

\subsection{The lifetime of unpolarized muons}

If the muons are unpolarized and if we do not observe the polarization of final-state
fermions, then the transition matrix element, Eq.~(\ref{eq:5}), is given by averaging
over the muon spin and summing over all final fermion spins:
\begin{eqnarray}\label{eq:27}
    M^2=&&\frac{G^2}{2}\;\frac{1}{2}\sum_{s,s',r,r'=1}^2\left[\;\overline{u}_{s'}(q)\;
    \gamma_\lambda\;(1+\gamma_5)\;v_{r'}(k')\;\right]^2\nonumber\\
    &&\times\left[\;\overline{u}_r(k)\;\gamma_\lambda\;(1+\gamma_5)\;u_s(p)\;\right]^2.
\end{eqnarray}
where $p$, $q$, $k$ and $k'$ are 4-momenta, while $s$, $s'$, $r$ and $r'$ are spin
indices for $\mu$, $e$, $\nu_\mu$ and $\bar{\nu}_e$, respectively. For the convenience of
discussion below, in Eq.~(\ref{eq:27}) we set
\begin{equation}\label{eq:28}
  I=\frac{1}{2}\sum_{s=1}^2\sum_{r=1}^2\left[\;\overline{u}_r(k)\;\gamma_\lambda\;
  (1+\gamma_5)\;u_s(p)\;\right]^2,
\end{equation}
which is related to the muons. By means of Eqs.~(\ref{eq:10}) and (\ref{eq:12}), the
evaluations of spin sums are reduced to the calculation of projection operators:
\begin{eqnarray}
  \sum_{s=1}^2u_s(p)\overline{u}_s(p)&=&\sum_{s=1}^2P_s(p)=\Lambda_+(p),\label{eq:29}\\
  \sum_{s=1}^2v_s(p)\overline{v}_s(p)&=&-\sum_{s=1}^2P_s(p)=-\Lambda_-(p).\label{eq:30}
\end{eqnarray}
One sees that the explicit evaluation of spin projection operators disappear. Applying
Eqs.~(\ref{eq:29}), (\ref{eq:30}) and (\ref{eq:11}) as well as the trace theorems we
obtain
\begin{equation}\label{eq:31}
  M^2=\frac{4\;G^2\;(p\cdot k')\;(q\cdot k)}{E_p E_q E_k E_{k'}}.
\end{equation}
Substituting Eq.~(\ref{eq:31}) into Eq.~(\ref{eq:26}), one has
\begin{equation}\label{eq:32}
  \tau^{-1}=\frac{4\;G^2}{(2\pi)^5}\frac{1}{E_p}\int\!\frac{d^3 q}{E_q}\frac{d^3 k}{E_{k}}\frac{d^3
  k'}{E_{k'}}\delta^4(p-q-k-k')\cal F,
\end{equation}
where
\begin{equation}\label{eq:33}
  {\cal F}=(p\cdot k')\;(q\cdot k).
\end{equation}
Obviously decay amplitude $\cal F$ is a Lorentz-invariant matrix element. Therefore we
have
\begin{equation}\label{eq:34}
  {\cal F}={\cal F}^0=(p^0\cdot k')(q\cdot k),\quad p^0=(0,0,0,im_\mu)
\end{equation}
where ${\cal F}^0$ is $\cal F$ in the muon rest frame.

It is easy to see from Eq.~(\ref{eq:32}) that the integration to the right of $E_p^{-1}$
is Lorentz invariant. Neglecting electron mass, the muon lifetime $\tau_{_0}$ in its rest
frame is given by
\begin{equation}\label{eq:42}
    \tau^{-1}_0=\frac{G^2m^5_\mu}{192\;\pi^3},
\end{equation}
where $m_\mu$ is muon mass. In an arbitrary frame the muon lifetime is given by
\begin{equation}\label{eq:43}
    \tau=\frac{\tau_{_0}}{\sqrt{1-\beta^2}}.
\end{equation}
We see that the lifetime is proportional to the velocity $\beta$ of muons as required by
special relativity$^{[12]}$. So that the lifetime is not a Lorentz scalar.

\subsection{The lifetime of polarized muons expressed by the spin states}

For polarized muons the muon spin should not be averaged. In most literatures and
textbooks$^{[2,13]}$ the polarized states of fermions were usually expressed by the spin
states (\ref{eq:6}) and (\ref{eq:7}). Instead of Eqs.~(\ref{eq:29}) and (\ref{eq:30}), we
have
\begin{equation}\label{eq:35}
  I_s=\sum_{r=1}^2\left[\overline{u}_r(k)\gamma_\lambda(1+\gamma_5)u_s(p)\right]^2,
\end{equation}
and
\begin{equation}\label{eq:36}
  u_s(p)\overline{u}_s(p)=P_s(p)=\frac{1}{2}(1\pm i\gamma_5\gamma\cdot e)\frac{(-i\gamma\cdot p+m_\mu)}{2E_p},
\end{equation}
respectively. Substituting Eq.~(\ref{eq:36}) into (\ref{eq:35}) and Eq.~(\ref{eq:35})
into (\ref{eq:28}) we obtain
\begin{equation}\label{eq:37}
  M_s^2=\frac{4\;G^2{\cal F}_s}{E_p E_q E_k E_{k'}},
\end{equation}
where
\begin{equation}\label{eq:38}
  {\cal F}_s=(p\cdot k')\;(q\cdot k)\mp m_\mu(e\cdot k')\;(q\cdot k)
  ={\cal F} \mp {\cal F}_e,
  \end{equation}
where ${\cal F}_e$ is the decay amplitude related to the polarization vector $e$:
\begin{equation}\label{eq:39}
  {\cal F}_e=m_\mu(e\cdot k')\;(q\cdot k).
\end{equation}
Obviously, decay amplitude ${\cal F}_s$ is also a Lorentz scalar, like $\cal F$.
Therefore we have
\begin{equation}\label{eq:40}
  {\cal F}_s={\cal F}^0_s=(p^0\cdot k')\;(q\cdot k)\mp m_\mu(e^0\cdot k')\;(q\cdot k),
\end{equation}
where ${\cal F}^0_s$ is ${\cal F}_s$ in the muon rest frame.

In a similar way to Eq.~(\ref{eq:32}), we obtain
\begin{eqnarray}\label{eq:41}
    \tau^{-1}_s&=&\frac{4\;G^2}{(2\pi)^5}\frac{1}{E_p}\int\!\frac{d^3 q}{E_q}\frac{d^3
    k}{E_{k}}\frac{d^3k'}{E_{k'}}\delta^4(p-q-k-k'){\cal F}_s\nonumber\\
    &=&\tau^{-1}+\tau^{-1}_e.
\end{eqnarray}
One easily verifies that the second integration over ${\cal F}_e$ vanishes, i.e.,
$\tau_e=0$. The first integration is identical with Eq.~(\ref{eq:32}) and we obtain
\begin{equation}\label{eq:41-1}
    \tau_s=\tau.
\end{equation}
It is also easy to see in the above calculation that the lifetime in the laboratory frame
does not exhibit any lifetime asymmetry when the polarization of muons is described by
spin states.

\subsection{The lifetime of polarized muons expressed by the helicity states}

As mentioned above, however, this method does not enable us to discuss the dependence of
lifetime on the polarization of muons. Because a spin state is helicity degenerate, the
polarization of muons must be described by helicity states. For LH polarized muons,
substituting the spin states in Eq.~(\ref{eq:35}) with the helicity states and
considering Eq.~(\ref{eq:15-1}), we obtain
\begin{eqnarray}
    I_{Lh}&=&\sum_{h}\left[\overline{u}_h(k)\gamma_\lambda
    (1+\gamma_5)u_{_{Lh}}(p)\right]^2\nonumber\\
    &=&\sum_{r=1}^2\left[\overline{u}_r(k)\gamma_\lambda
    (1+\gamma_5)u_{_{Lh}}(p)\right]^2.\label{eq:44}
\end{eqnarray}
From Eqs.~(\ref{eq:21}), (\ref{eq:24}) and (\ref{eq:18}) we easily find
\begin{equation}\label{eq:45}
  (1+\gamma_5)u_{_{Lh}}(p)=2\sqrt{1+\beta}u^0_{_{L2}}=\sqrt{1+\beta}(1+\gamma_5)u^0_2,
\end{equation}
where $u^0_2$ is the spin state in the muon rest frame. One can see that the
chirality-state projection operator $(1+\gamma_5)$ picks out LH chirality state
$u^0_{_{L2}}$ in a LH helicity state, which is factorized into two parts in the second
equation, one is the spin state $u^0_2$ and another is a factor $\sqrt{1+\beta}$ which
depends on muon's helicity. Substituting Eq.~(\ref{eq:45}) into Eq.~(\ref{eq:44}) we have
\begin{equation}\label{eq:46}
    I_{Lh}=(1+\beta)\sum_{r=1}^2\left[\overline{u}_r(k)\gamma_\lambda
    (1+\gamma_5)u^0_2\right]^2
\end{equation}
Comparing Eq.~(\ref{eq:46}) with Eq.~(\ref{eq:35}) and considering Eq.~(\ref{eq:40}) we
find out the decay amplitude of LH polarized muons
\begin{equation}\label{eq:47}
  {\cal F}_{Lh}=(1+\beta){\cal F}_2^0=(1+\beta){\cal F}_2.
\end{equation}
Then the LH polarized fermion lifetime is given by
\begin{equation}\label{eq:48}
    \tau^{-1}_{_{Lh}}=\frac{4\;G^2}{(2\pi)^5}\frac{1}{E_p}\int\!\frac{d^3 q}{E_q}\frac{d^3
    k}{E_{k}}\frac{d^3k'}{E_{k'}}\delta^4(p-q-k-k'){\cal F}_{Lh}.
\end{equation}

Similarly, for RH polarized muons we obtain
\begin{eqnarray}\label{eq:49}
    I_{Rh}&=&\sum_{h}\left[\overline{u}_h(k)\gamma_\lambda
    (1+\gamma_5)u_{_{Rh}}(p)\right]^2\nonumber\\
    &=&(1-\beta)\sum_{r=1}^2\left[\overline{u}_r(k)\gamma_\lambda
    (1+\gamma_5)u^0_1\right]^2
\end{eqnarray}
and the decay amplitude is
\begin{equation}\label{eq:50}
    {\cal F}_{Rh}=(1-\beta){\cal F}_1^0=(1-\beta){\cal F}_1.
\end{equation}
Then the RH polarized fermion lifetime is given by
\begin{equation}\label{eq:51}
    \tau^{-1}_{_{Rh}}=\frac{4\;G^2}{(2\pi)^5}\frac{1}{E_p}\int\!\frac{d^3 q}{E_q}\frac{d^3
    k}{E_{k}}\frac{d^3k'}{E_{k'}}\delta^4(p-q-k-k'){\cal F}_{Rh}.
\end{equation}

Comparing Eqs.~(\ref{eq:47}) and (\ref{eq:50}) with Eq.~(\ref{eq:38}), respectively and
considering Eqs.~(\ref{eq:41}), (\ref{eq:41-1}) and (\ref{eq:43}), we find the polarized
muon lifetime
\begin{equation}\label{eq:52}
    \tau_{_{Lh}}=\frac{\tau}{1+\beta},\quad\hbox{and}\quad
    \tau_{_{Rh}}=\frac{\tau}{1-\beta}.
\end{equation}

\begin{figure}[htb!]
\includegraphics{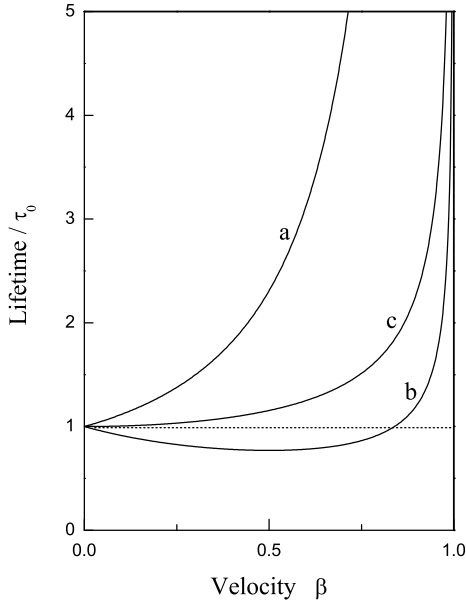}
\caption{\label{fig:lifetime}Lifetime as a function of muon velocity $\beta$. (a) The
lifetime $\tau_{_{Rh}}$ of right-handed polarized muons. (b) The lifetime $\tau_{_{Lh}}$
of left-handed polarized muons. (c) The lifetime $\tau$ of unpolarized muons.}
\end{figure}

It is easy to see that the $\tau_{_{Rh}}$ is greater than $\tau_{_{Lh}}$, which shows the
lifetime asymmetry of left-right handed polarized fermions. The lifetime asymmetry is
expressed by
\begin{equation}\label{eq:53}
    \hbox{lifetime asymmetry}\equiv\frac{\tau_{_{Rh}}-\tau_{_{Lh}}}{\tau_{_{Rh}}+\tau_{_{Lh}}}
    =\beta.
\end{equation}
When $\beta=0$, we find $\tau_{_{Rh}}=\tau_{_{Lh}}=\tau_{_0}$; when $\beta\rightarrow 1$,
we find $\tau_{_{Rh}}\rightarrow\infty$, $\tau_{_{Lh}}\rightarrow\infty$. In particular,
when $\beta=\frac{1}{2}$, the lifetime of the LH polarized fermions has a minimum value
$\tau_{_{Lh}}=\tau_{min}=0.77\tau_{_0}$, as shown in Fig.~\ref{fig:lifetime}.

It is not difficult to prove that the lifetimes of antimuons $\mu^+$ in flight are given
by
\begin{equation}\label{eq:54}
    \overline\tau_{_{Rh}}=\frac{\tau}{1+\beta},\quad\hbox{and}\quad
    \overline\tau_{_{Lh}}=\frac{\tau}{1-\beta}.
\end{equation}

\section{summary and discussion}

The calculation has established that the lifetime of RH polarized muons is different from
that of LH polarized muons in flight. Under a space reflection the LH helicity state
Eq.~(\ref{eq:21}) and the RH helicity state Eq.~(\ref{eq:22}) transform to each other,
therefore the set of equations Eq.~(\ref{eq:52}) and Eq.~(\ref{eq:54}) are valid all the
same, respectively. Furthermore, this conclusion is also valid for all fermions in the
decays under weak interactions. It means that the $\tau_{_{Lh}}$ is always smaller than
$\tau_{_{Rh}}$ for fermions and the $\overline \tau_{_{Rh}}$ smaller than
$\overline\tau_{_{Lh}}$ for antifermions in any one of inertial systems in which fermions
or antifermions are in flight with a same speed. The lifetime asymmetry shows a maximum
violation of parity symmetry. Hence under the condition of parity violation the lifetime
is neither a four-dimensional scalar, nor a scalar under the three-dimensional space
inversion.

We emphasize here an important concept that a spin state is helicity degenerate and the
spin projection operators $\rho_{\pm}$ can only project out the spin states, but can not
project out the helicity states. The above calculation shows that the so-called
eigenstate of operator given by Eq.~(\ref{eq:8}) is by no means a helicity eigenstate
(even we had chosen $\bm p$ vector along z axis) and this is why the parity violation
result, Eqs.~(\ref{eq:52}) and (\ref{eq:54}), was overlooked in the past for so long a
time even one did not perform the spin average for muons in the laboratory frame.
Therefore, the polarized fermions must be expressed by the helicity states which are
relevant to physical interpretation and experimental test.

If we find out nonzero pseudoscalar of similar $\langle\bm \sigma\cdot\bm p\rangle$ in a
physical process, then parity conservation law is certainly violated and Lorentz
invariance is also violated because $\langle\bm \sigma\cdot\bm p\rangle$ is not a Lorentz
scalar. In the special theory of relativity space coordinates are treated on an equal
footing with time coordinate. The violation of pure space inversion, in some sense,
implies a violation to the equal status and mutual transformation between space and time.
And it in turn implies that something might go beyond the theory of special
relativity$^{[14]}$. The lifetime asymmetry is just such an example. It poses a serious
challenge to beautiful symmetry in physics.

%The pure space inversion violates the equalization between space and time and thus parity
%nonconservation certainly leads to some physics beyond the special relativity$^{[14]}$.
%The lifetime asymmetry is just a good example of such a physics.

When the polarization of muons is described by spin states, it can be seen from
Eq.~(\ref{eq:39}) that the factor related to polarization reads
\begin{equation}
    (e\cdot k')=(e\cdot p)-(e\cdot q)-(e\cdot k).
\end{equation}
The pseudoscalars $\bm e\cdot \bm q$ and $\bm e\cdot \bm k$ are included in it, which
reveals the asymmetric distribution of electrons and neutrinos emitted from decaying
muons and exhibits parity nonconservation in the final channels. The property $(e\cdot
p)=0$ ensures the pseudoscalar $\bm e\cdot\bm p$, helicity operator, does not appear in
the decay amplitude. Therefore, the decay amplitude corresponding to the spin states is
independent of muon helicity, which is consistent with the helicity degeneracy of the
spin states.

Differing from the spin states, however, for massive muons
%an observer moving with a
%velocity, which parallels and is greater than the muon's velocity, would see a LH
%polarized muon becoming a RH one or vice versa. That is to say,
LH and RH helicity state may transform to each other as observer's velocity paralleling
and exceeding the muon's one. If, therefore, the decay amplitude corresponding to the
helicity state were a Lorentz scalar, then the decay amplitude of LH polarized muons
would be equal to that of RH one and starting from Eqs.~(\ref{eq:47}), (\ref{eq:50}) and
(\ref{eq:38}), we would obtain ${\cal F}_{Lh}={\cal F}_{Rh}=(p\cdot k')(q\cdot k).$
Obviously, there exists no any pseudoscalars in it, which shows that the parity is
conserved and is clearly wrong. Hence we conclude that the decay amplitude ${\cal
F}_{Lh}$ and ${\cal F}_{Rh}$ corresponding to the helicity states must be not Lorentz
invariant, as shown Eqs.~(\ref{eq:47}) and (\ref{eq:50}). It is the root cause of the
lifetime asymmetry. In other words, if the polarization is described by the helicity
states, then the lifetime asymmetry will be inevitable. Of course, the average value of
decay amplitude, $\frac{1}{2}({\cal F}_{Lh}+{\cal F}_{Rh})$, is still a Lorentz scalar.

In the before, physicists too believed in the covariant form of four spinor which enjoys
the invariance under the LT and so misunderstood the meaning of so-called spin states .
Meanwhile, the importance of helicity state and its difference from spin state and chiral
state were often confused or overlooked. Our lesson and experience are just focused on
the above crucial point before we are able to get rid of the constraint of covariant four
spinor form. What we eventually realized is to discriminate two things: while a physical
law like that reflected by the Dirac equation is one thing, a phenomenon like a
particle's velocity, its helicity as well as its decay lifetime is another thing. A law
should be invariant under the LT , but a phenomenon needs not. Actually, a phenomenon may
be different in different inertial systems. In short, the information is not existing
totally in the sense of objectiveness, it is created by the subject and object in common.

The measurements on muon decay used to be performed in its rest frame. It was realized
that muons, formed by forward decay in flight of pions inside cyclotron, were stopped in
a nuclear emulsion, sulphur, carbon, calcium or polyethylene target. The polarization
effects of muon decay were observed using carbon stopping target$^{[15]}$, in which there
is no depolarization of the muons. So far the measurement of the lifetime of polarized
muons in flight has not yet been found in literature. Therefore, one actually lacks any
direct experimental evidence either to support or to refute the lifetime asymmetry. We
report it here now in the hope that it may stimulate and encourage further experimental
investigations on the question of the lifetime asymmetry in muon decays.

%\begin{acknowledgments}
%We are grateful to Dr. T. Yamashita and Dr. LiMing for their many helpful discussions.
%\end{acknowledgments}

\end{document}